\documentclass[twoside,twocolumn,a4paper]{article}
\usepackage{graphicx}
\usepackage{blindtext}
\usepackage{lettrine}

\usepackage{bm}
\usepackage{amsmath}
\usepackage{epstopdf}
\usepackage{hyperref}

\usepackage{epstopdf, epsfig}
\usepackage{textcomp}
\usepackage{color}

\usepackage[sc]{mathpazo}
\usepackage[T1]{fontenc} 
\usepackage{microtype} 

\usepackage[english]{babel} 

\usepackage[hmarginratio=1:1,top=20mm,columnsep=15pt,left=18mm]{geometry} 
\usepackage[hang, small,labelfont=bf,up,textfont=it,up]{caption} 
\usepackage{booktabs} 

\usepackage{lettrine} 

\usepackage{enumitem} 
\setlist[itemize]{noitemsep} 

\usepackage{abstract}

\usepackage{titlesec} 
\titleformat{\section}[block]{\large\scshape\centering}{\thesection.}{1em}{} 
\titleformat{\subsection}[block]{\large}{\thesubsection.}{1em}{} 

\usepackage{fancyhdr} 
\usepackage{titling} 
\usepackage{hyperref} 
\usepackage{authblk}

\pretitle{\begin{center}\Huge\bfseries} 
\posttitle{\end{center}} 
\title{Inverse centrifugal effect induced by collective motion of vortices in rotating turbulent convection} 
\author[1]{Shan-Shan Ding}
\author[2]{Kai Leong Chong}
\author[1]{Jun-Qiang Shi}
\author[3,2]{Guang-Yu Ding}
\author[1]{Hao-Yuan Lu}
\author[3,2]{Ke-Qing Xia\thanks{xiakq@sustech.edu.cn}}
\author[1]{Jin-Qiang Zhong\thanks{jinqiang@tongji.edu.cn}}
\affil[1]{School of Physics Science and Engineering, Tongji University, Shanghai 200092, China.}
\affil[2]{Department of Physics, The Chinese University of Hong Kong, Shatin, Hong Kong, China.}
\affil[3]{Center for Complex Flows and Soft Matter Research and Department of Mechanics and Aerospace Engineering, Southern University of Science and Technology, Shenzhen 518055, China.}
\date{\today}
\renewcommand{\maketitlehookd}{%

\begin{abstract}
\noindent 
When a fluid system is subject to strong rotation, centrifugal fluid motion is expected, i.e., denser (lighter) fluid moves outward (inward) from (toward) the axis of rotation. 
Here we demonstrate, both experimentally and numerically, the existence of an unexpected outward motion of warm and lighter vortices in rotating turbulent convection. This anomalous vortex motion occurs under rapid rotations when the centrifugal buoyancy is sufficiently strong to induce a symmetry-breaking in the vorticity field, i.e., the vorticity of the cold anticyclones overrides that of the warm cyclones. We show that through hydrodynamic interactions the densely populated vortices can self-aggregate into coherent clusters and exhibit collective motion in this flow regime. Interestingly, the correlation of the vortex velocity fluctuations within a cluster is scale-free, with the correlation length being about $30\%$ of the cluster length. Such long-range correlation leads to the collective outward motion of cyclones. Our study provides new understanding of vortex dynamics that are widely present in nature.
\end{abstract}
}

\begin{document}

\maketitle

{\lettrine{C}{}oherent vortex structures exist ubiquitously in many flow systems ranging from small-scale turbulence to large-scale geophysical and astrophysical flows \cite{Va06, Mc06, HV93}, and their dynamics plays a crucial role in determining turbulent mixing and transport in those systems.
Previous studies of vortex dynamics are mainly focused on isolated vortices \cite{HV93, VK89}. However, densely populated vortices are observed in rapidly rotating turbulence \cite{AMetal18, YB20}, and the resulting vortex interactions may lead to markedly different dynamics compared to that of isolated vortices \cite{LM93, BW99, FS01}. Many nonequilibrium statistical systems in nature consisting of densely populated,  interacting entities often exhibit collective behavior, i.e., the entities self-aggregate to perform collective motions. Examples include flocking of bird, swarming of bacteria and clustering of active matters, etc \cite{CCGPSSV10, CDBSZ12, KFSH17}. Whether collective behavior of vortices can arise in rotating turbulent flows is thus a question of fundamental interest.


Here we demonstrate both experimentally and numerically the collective motion of vortices in rotating turbulent convection. The resultant long-range correlated vortex dynamics gives rise to the counterintuitive effect of inverse centrifugal motion, i.e., the warm and lighter convective vortices exhibit outward motion from the rotation axis. This intriguing phenomenon occurs in a rapidly rotating regime where the strong centrifugal buoyancy breaks the symmetry in both the population and vorticity magnitude of the vortices. Our study reveals that it is through local hydrodynamic interactions that the densely populated vortices self-aggregate into large-scale vortex clusters, in which the warm cyclonic vortices submit to the collective motion dominated by the strong anticyclones and move outwardly.


\begin{figure*}
	\includegraphics[width=1.0\textwidth]{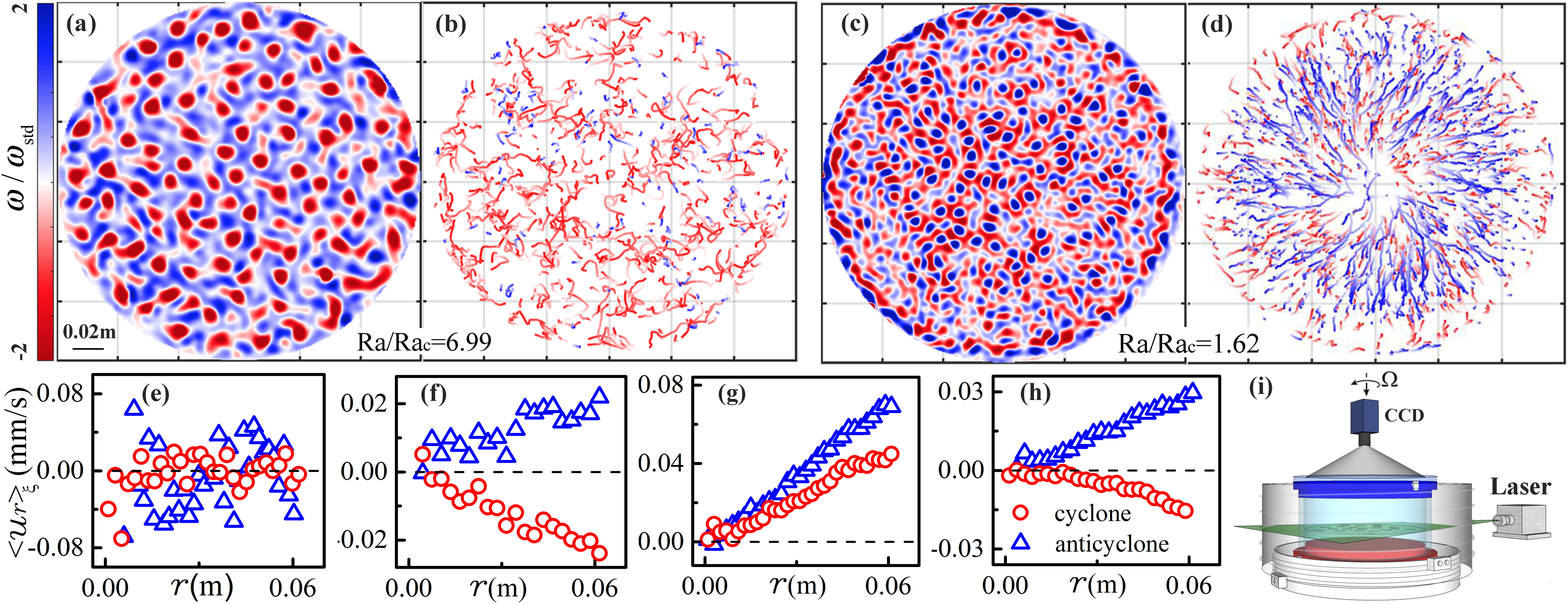}
\caption{(a, c) Instantaneous vertical vorticity distribution $\omega/{\omega}_{std}$ over the measured fluid height. ${\omega}_{std}$ is the standard deviation of $\omega$. (b, d) Trajectories for cyclones (red) and anticyclones (blue). The shading of the trajectories indicates that the vortices appear (terminate) at the light (dark)-color side. Results for $\mathrm{Ra}{=}3.0{\times}10^{7}$ and $\mathrm{Ra/Ra_c}{=}6.99$, Fr${=}0.03$ (a, b);  $\mathrm{Ra/Ra_c}{=}1.62$, Fr${=}0.27$ (c, d). (e-h) Radial profiles $\langle{u}_r(r)\rangle_{\xi}$ of cyclones and anticyclones in the four flow regimes for Ra${=}2.0{\times}10^7$ and from left to right, $\mathrm{Ra/Ra_{c}}{=}15.6, 3.61,1.83,1.17$. ${\langle}...{\rangle}_{\xi}$ denotes a trajectory-ensemble average. (i) Schematic of the experimental set-up. A laser sheet illuminates a rotating Rayleigh-B\'enard convection cell filled with water and seeded with tracer particles at a fluid height $z{=}H/4$. A co-rotating camera images the light scattered by the tracer particles.}
\label{fig1}
\end{figure*}

The fluid dynamics of rotating turbulent flows is often studied in rotating Rayleigh-B\'enard convection (RBC). Despite the considerable progress achieved in studying non-rotating and weakly rotating turbulent RBC \cite{AGL09, LX10, Xi13, LE97, KCG06, ZSCVLA09},
some important convection regimes that may exhibit novel vortex dynamics are yet to be explored (see, e.g. \cite{HA18, Paper1}).
Recent studies report that in rapidly rotating RBC, the convection flows are organized by the Coriolis force into coherent columnar vortices \cite{BG86, Sa97, VE02, PKVM08, KSNHA09, GJWK10, KCG10, JRGK12, KA12, SLDZ20}. 
These columnar vortices are helical structures with their vorticity correlated to the temperature, i.e., cyclones are warm (upwelling) vortices while anticyclones are cold (downwelling) ones if observed from near the bottom boundary \cite{Sa97, KA12}. The Taylor-Proudman theorem \cite{Tr88}, which predicts that rapid rotation suppresses flow variations along the vertical axis of rotation, provides a comprehensive description of the flow structures of the columnar vortices. From a dynamical viewpoint, however, the horizontal motion of, and the interaction between these vortices remain to be better understood.




\begin{figure*}
	\centering
	\setlength{\unitlength}{\textwidth}
\includegraphics[width=0.85\textwidth]{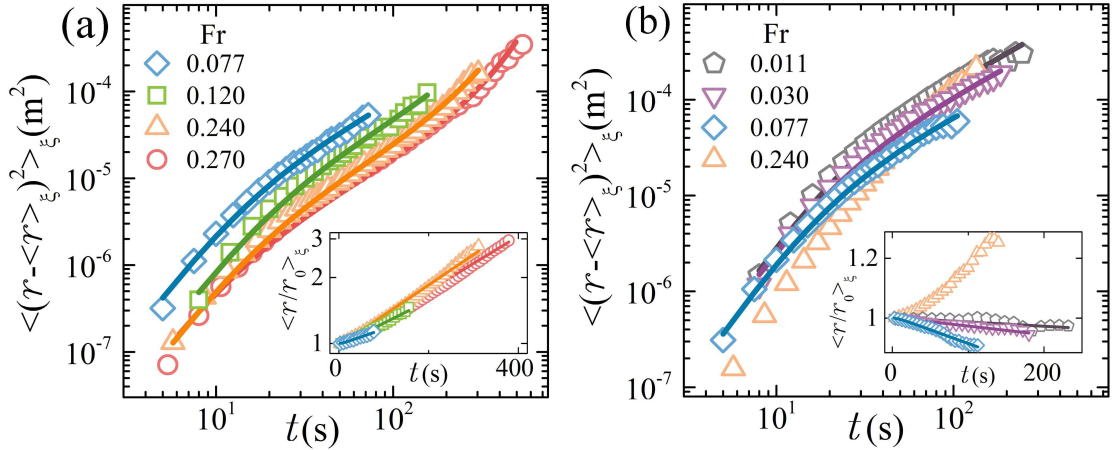}
\caption{Experimental data (open symbols) and theoretical predictions (solid lines) of the second moment of radial displacement ${\langle}(r{-}{\langle}r{\rangle}_{\xi})^{2}{\rangle}_{\xi}$ for anticyclones (a) and cyclones (b). 
Results for Ra${=}3.0{\times}10^7$. The insets show the results of the first moment of radial displacement ${\langle}r/r_0{\rangle}_{\xi}$. $r_0$ is the initial radial position of a vortex.}
\label{fig:2}
\end{figure*}

Our experimental apparatus was designed for high-precision flow structure measurement in rotating RBC \cite{Paper1, SLDZ20}. 
We used cylindrical cells with an inner diameter $d{=}$240 mm and length $H{=}$63.0 (120.0) mm, yielding the aspect ratio $\Gamma{=}d/H{=}$3.8 (2.0). The experiment was conducted with a constant Prandtl number Pr${=}\nu/{\kappa}{=}4.38$ and in the range $2.0{\times}10^7{\le}$Ra${\le}2.7{\times}10^8$ of the Rayleigh number Ra${=}{\alpha}g{\Delta}TH^3/{\kappa}{\nu}$ ($\alpha$ is the isobaric thermal expansion coefficient, $g$ the acceleration of gravity, $\Delta T$ the applied temperature difference, $\kappa$ the thermal diffusivity, and $\nu$ the kinematic viscosity). Rotation rates up to 5.0 rad/s were used. Thus the Ekman number Ek${=}\nu/2{\Omega}H^2$ spanned $4.9{\times}10^{-6}{\le}$Ek${\le}2.7{\times}10^{-4}$, corresponding to a range of the reduced Rayleigh number $1.06{\le}\mathrm{Ra/Ra_c}{\le}120$, with $\mathrm{Ra_c}{=}8.7$Ek$^{-4/3}$ the critical value for the onset of convection \cite{Ch61}. The Froude number Fr${=}{\Omega}^2d/2g$ was within $0{<}$Fr${\le}0.31$. The flow field at a fluid depth of $z{=}H/4$ was measured using the technique of particle image velocimetry (PIV) (See schematic of the experimental set-up in Fig.\ 1i). 
In the direct numerical simulation (DNS) we solved the Navier-Stokes equations with the Coriolis and centrifugal forces included, using the multiple-resolution version of the {\it{CUPS}} code \cite{KX13, CDX18}. The simulation was performed in a cylindrical domain with $\Gamma{=}4$, $\mathrm{Ra}{=}2.0{\times}10^{7}$ and $1.06{\le}\mathrm{Ra/Ra_c}{\le}40$.


Figure 1 presents snapshots of the vortex structures over the measured fluid height ($z=H/4$). At a low rotation rate when the centrifugal force is negligible (Fig.\ 1a), the cyclonic vortices (shown in red color) possess a greater number density and on average larger vorticity in magnitude than that of anticyclones (blue color). Both types of vortices exhibit stochastic horizontal motions as indicated by the vortex trajectories in Fig.\ 1b, presumably due to the turbulent background flows. The mean-square-displacement (MSD) of the vortices becomes a linear function of time at large times, indicating a Brownian-type, normal diffusive motion \cite{Paper1}.

However, at higher rotation rates when the centrifugal force becomes dominant, we observe strong anticyclones with larger population than the cyclones (Fig.\ 1c). The anticyclones undergo outward radial motions accompanied by stochastic fluctuations along their paths (Fig.\ 1d), until they move close to the sidewall where their radial motion is terminated by the retrogradely traveling plumes. Compared to the anticyclones, the motion of weak cyclones are much more complex. Figure 1d indicates that in the outer region ($r{\ge}d/4$), the cyclones move toward the cell center while in the inner region ($r{\le}d/4$) most of them migrate radially outward. In this rapid rotating case, the MSD of both types of vortices indicate superdiffusive behavior (see Supplementary Movies of the vortex motions).

\begin{figure*}
	\centering
	\setlength{\unitlength}{\textwidth}
	\includegraphics[width=1.0\textwidth]{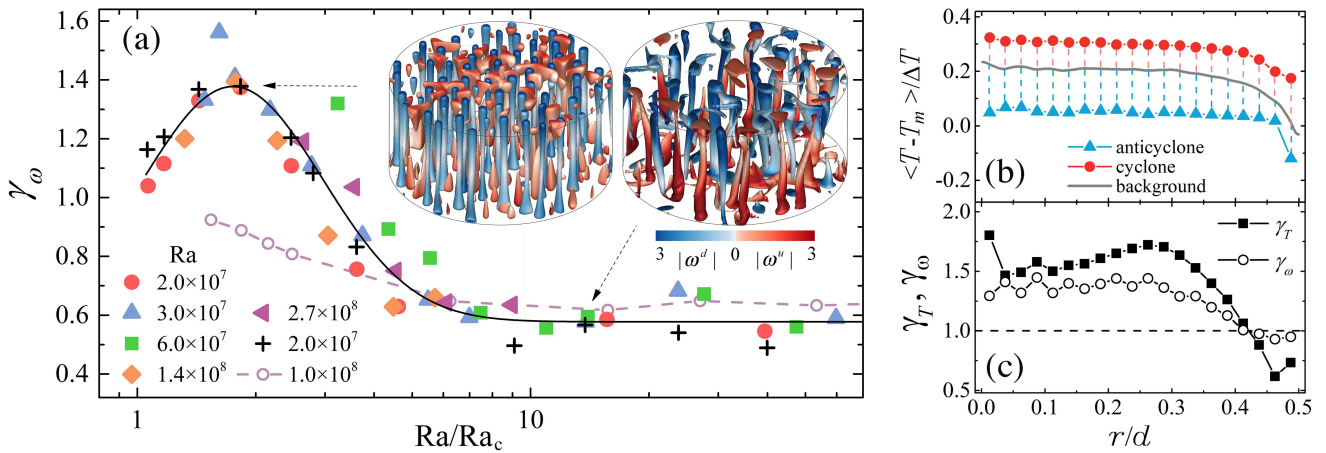}
\caption{(a) The vorticity ratio $\gamma_{\omega}{=}{\lvert}\langle{\omega}_{a}\rangle/\langle{\omega}_{c}\rangle{\rvert}$ of the anticyclones to the cyclones as a function of $\mathrm{Ra/Ra_{c}}$. Here ${\langle}...{\rangle}$ denotes a time average. Filled symbols: experimental data for $\Gamma{=}3.8$ with Ra${=}2.0{\times}10^7$ (circles), $3.0{\times}10^7$ (up triangles), $6.0{\times}10^7$ (squares); and for $\Gamma{=}2.0$ with Ra${=}1.4{\times}10^8$ (diamonds), $3.0{\times}10^7$ (left triangles). 
Data from DNS including (excluding) the centrifugal force are shown in pluses (open circles) for $\Gamma{=}4.0$. The solid curve indicates the trend of the experimental data. Inset panels: iso-surfaces of the temperature anomaly from DNS for Ra${=}2.0{\times}10^7$, $\mathrm{Ra/Ra_{c}}{=}1.83$ (left) and $\mathrm{Ra/Ra_{c}}{=}13.76$ (right). The coloration represents the vorticity magnitude of the uprising ($\omega^u$) and downwelling ($\omega^d$) vortices. (b) Radial profiles of the mean temperature ${\langle}T{-}T_m{\rangle}/{\Delta T}$ for the vortices and the background fluid. $T_m$ is the mean of the top and bottom fluid temperature. The length of the dashed lines indicates the temperature difference $\delta{T}$ between the cyclones (anticyclones) and the background fluid. (c) Radial profiles of ${\gamma}_{\omega}$ and ${\gamma}_{T}{=}\lvert{\langle}{\delta T_{a}}{\rangle}/{\langle}{\delta T_{c}}{\rangle}\rvert$. (b) and (c) are DNS data for Ra${=}2.0{\times}10^7$ and $\mathrm{Ra/Ra_{c}}{=}1.83$.}
\label{fig3}
\end{figure*}


To further quantify the vortex motions, we show the profiles of mean radial velocity  ${\langle}u_r{\rangle}_{\xi}$ of the vortices measured in the inner region of the cell in Figs.\ 1e-1h. These velocity profiles reveal four distinct flow regimes depending on the rotation rates: 
(I) A randomly-diffusive regime exists in the slow rotating limit with Ra one order in magnitude larger than $\mathrm{Ra_c}$. In this regime the vortices move in a random manner, yielding ${\langle}u_r{\rangle}_{\xi}{\approx}0$ (Fig.\ 1e). 
(II) Centrifugation-influenced regime where the magnitude of ${\langle}u_r{\rangle}_{\xi}$ increases linearly with $r$ ($3\mathrm{Ra_c}{\le}\mathrm{Ra}{\le}5\mathrm{Ra_c}$). We observe that warm cyclones (cold anticyclones) move radially inward (outward), which is in agreement with the centrifugal effect (Fig.\ 1f). 
(III) Inverse-centrifugal regime ($1.5\mathrm{Ra_c}{\le}\mathrm{Ra}{\le}3\mathrm{Ra_c}$) in which there is anomalous outward cyclonic motion (Fig.\ 1g), and the radial gradients of ${\langle}u_r{\rangle}_{\xi}$ for both types of vortices reach a maximum. 
(IV) Asymptotic regime in the rapid rotation limit ($\mathrm{Ra}{\le}1.5\mathrm{Ra_c}$) where the opposite radial motions of cyclones and anticyclones recover (Fig.\ 1h). 

We formulate a theoretical model that consists a sets of Langevin equations incorporating the centrifugal force, which governs the radial vortex motion in a background of stochastic fluctuations. As shown in Fig.\ 2, the model provides predictions of the first and second moments of the radial vortex displacements which replicate very well the experimental data in flow regimes (II) and (IV) (see Supplementary Information 2 for detailed discussions of the model). Nonetheless, Figure 2 also suggests that the inverse centrifugal motion of the cyclones in the anomalous regime (III) remains unexplained on the basis of the model. A key question remained to be answered is then what sets the anomalous vortex motion?



To gain insights into the observed phenomenon, we first examine the relative strength of vorticity between the cyclones and anticyclones. Figure 3a shows the vorticity ratio $\gamma_{\omega}$ of the anticyclones to the cyclones. We note that in the randomly-diffusive regime, $\gamma_{\omega}$ is approximately 0.6, meaning that at the measured fluid height $z{=}H/4$ the cyclonic vorticity is overall larger in magnitude than the anticyclonic ones (see Fig.\ 1a). It is the case because in this regime the downwelling anticyclones generated from the top travel a longer distance to the measured fluid layer than the upwelling cyclones, with their momentum and vorticity largely dissipated by the background turbulence \cite{KCG10}. (The vorticity magnitude of the two types of vortices are equal if measured at $z{=}H/2$. See the visualization from our DNS in the right inset of Fig.\ 3a.) 
With increasing $\Omega$ the up- and down-welling vortices evolve into the vertically symmetric, columnar structures \cite{KSNHA09, GJWK10}, thanks to the Taylor-Proudman effect \cite{Tr88}. Hence one would expect that the vorticity strength of the cyclones and anticyclones become comparable, i.e., $\gamma_{\omega}$ approaches unity. 
Our DNS data indeed show this trend when the centrifugal force is switched off.
However, when centrifugal effect is dominant, both the experimental and DNS results reveal that $\gamma_{\omega}$ exceed unity considerably in the inverse-centrifugal regime, indicating an asymmetric vorticity field dominated by the anticyclones (left inset of Fig.\ 3a). In the asymptotic regime where the severe rotational constraint finally weakens the convective vortices, $\gamma_{\omega}$ eventually returns to unity, and the symmetry of the cyclonic and anticyclonic vorticity restores. Remarkably, we observe that the trend of $\gamma_{\omega}(Ra/Ra_c)$ is universal and independent of $Ra$ and $\Gamma$.


We demonstrate that the asymmetry of the vorticity field in the anomalous regime results from the centrifugal effect. Figure 3b presents the radial profiles of the mean temperature of the vortices and of the background fluid. These numerical data indicate a noticeable warming of the background fluid in the inner region, owing to the centrifugal effect \cite{HO99, LE11, HA19}. As a result, the temperature difference ${\delta}T$ of the cold anticyclones from the background exceeds that of the warm cyclones. Since ${\delta}T$ is proportional to the buoyancy forcing on the vortices, it is predicted to be positively correlated to the vorticity $\omega$ in recent theoretical models \cite{PKVM08, GJWK10}. Indeed Fig. 3c shows that ${\delta}T$ and $\omega$ are both larger in magnitudes for anticyclones than for cyclones in the inner region, which explains the asymmetry of the vorticity field ($\gamma_{\omega}${>}1).

\begin{figure*}
	\centering
	\setlength{\unitlength}{\textwidth}
	\includegraphics[width=0.56\textwidth]{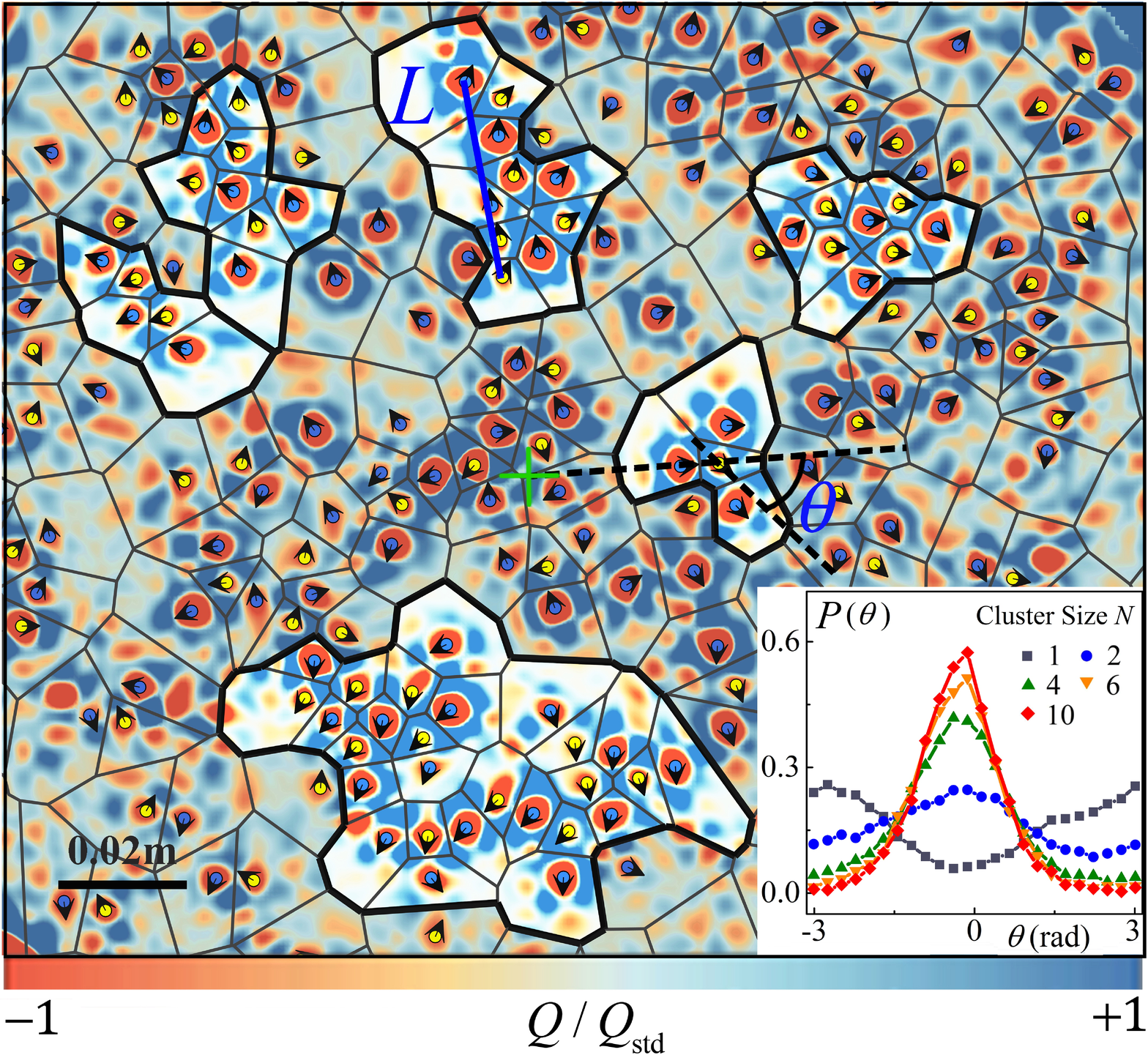}
\caption{Snapshot of the instantaneous velocity of the vortices in the central region for $\mathrm{Ra}{=}3.0{\times}10^7$, $\mathrm{Ra}/\mathrm{Ra_c}{=}1.62$, Fr${=}0.27$. Blue (yellow) circles show the centers of anticyclones (cyclones). Black arrows show the vortex velocity direction. The solid-line network represents the Voronoi diagram of the vortex centers. Examples of six vortex clusters are highlighted and marked with thick boundaries. $\theta$ denotes the angle between the position vector of a cyclone relative to the rotation axis (green cross) and its velocity. $L$ is the largest distance between two vortices within a cluster. The background coloration represents distribution of the quantity $Q/Q_{std}$ that reveals the strength of the vorticity field. Inset: Probability density functions of $\theta$ of cyclones in clusters with various size $N$.}
\label{fig4}
\end{figure*}

In the background of the broken symmetry of the vorticity field in the anomalous regime, we show here that it is the long-range correlated vortex motion that gives rise to the inverse centrifugal motion of the cyclones. Figure 4 presents the instantaneous motion of the vortices, with their spatial distribution presented in a Voronoi diagram. One sees that the adjacent vortices often self-organize into vortex clusters, i.e., the vortices move largely in the same direction. We adopt two criteria to identify vortex clusters, i.e., the distance of two neighboring vortices is smaller than 1.5 times the mean vortex diameter and the angle $\phi$ between their velocity vectors is within a threshold ($\phi{\le}\phi^{\ast}{=}60^{\circ}$). Our analysis over the range $30^{\circ}{\le}\phi^{\ast}{\le}75^{\circ}$ confirms that the results of correlated vortex motion are not sensitive to the choice of $\phi^{\ast}$. The direction of the motion of each cyclone $i$ is represented by the angle $\theta$ between its position vector $\vec{r}_i$ relative to the rotation axis and its velocity $\vec{u}_i$. We find that $\theta$ is strongly dependent on the number ($N$) of vortices in a cluster (inset). For isolated cyclones ($N{=}1$), the most probable direction of motion is radially inward ($\theta_p{=}\pi$). 
However, for clustered cyclones ($N{>}1$) we find $\theta_p{=}0$ as they move outward. The standard deviation of $\theta$ decreases monotonically when $N$ increases. Our data reveal that within large clusters the motion of weak cyclones submit to that of strong anticyclones and move outwardly in a collective manner. Their inverse centrifugal motion becomes more unidirectional with the increasing of the cluster size.


Some insights into the physical mechanism responsible for the vortex-cluster formation can be gained through statistical analysis of the cluster size distribution $p(N)$. Figure 5a shows that $p(N)$ for various $Ra/Ra_c$ can be well described by $p(N){=}AN^{-b}e^{-N/N_c}$. Here $b$ and $N_c$ are the fitting parameters with their dependence of $Ra/Ra_c$ plotted in the inset. For clusters in small size $N$, $p(N)$ first decays as a power function $N^{-b}$ up to a cutoff size $N_c$.  
Studies of the collective behavior in various natural systems have revealed that local aggregation of interacting entities is the essential ingredient for the power-law decay of the group-size distributions in these systems \cite{TNT88, Ta89, BDF99}. 
In the present vortex system, each vortex is surrounded more likely by counter-rotating vortices in a densely populated state (Fig.\ 4). 
Owing to the vortex-pair interaction, adjacent vortices of opposite-sign tend to move in similar directions \cite{LLW16} (see Supplementary Information 3 for details). 
Moreover, the shield structure forming near the edge of each vortex prevents strong interactions in closer proximity \cite{GJWK10, JRGK12, SLDZ20}, such as vortex merging and annihilation. As a result, isolated vortices are often aggregated into neighboring clusters and move collectively.
The power-law exponent, $b{=}1.5{\pm}0.04$, is found to be independent of $Ra/Ra_c$ (inset of Fig.\ 5a), and falls into the range of previous theoretical predictions \cite{TNT88, BD95}. For large $N$ we find that $p(N)$ evolves into an exponential tail with the cutoff size $N_c$ varying with $Ra/Ra_c$ and reaching a maximum at $Ra/Ra_c{=}1.62$, where the vorticity ratio $\gamma_{\omega}$ is maximum (see Fig.\ 3a). The rescaled data $p(N)N_c^{1.5}/A$ as a function of $N/N_c$ thus collapse into one master curve as shown in Figure 5b. 

\begin{figure*}
	\centering
	\setlength{\unitlength}{\textwidth}
	\includegraphics[width=0.70\textwidth]{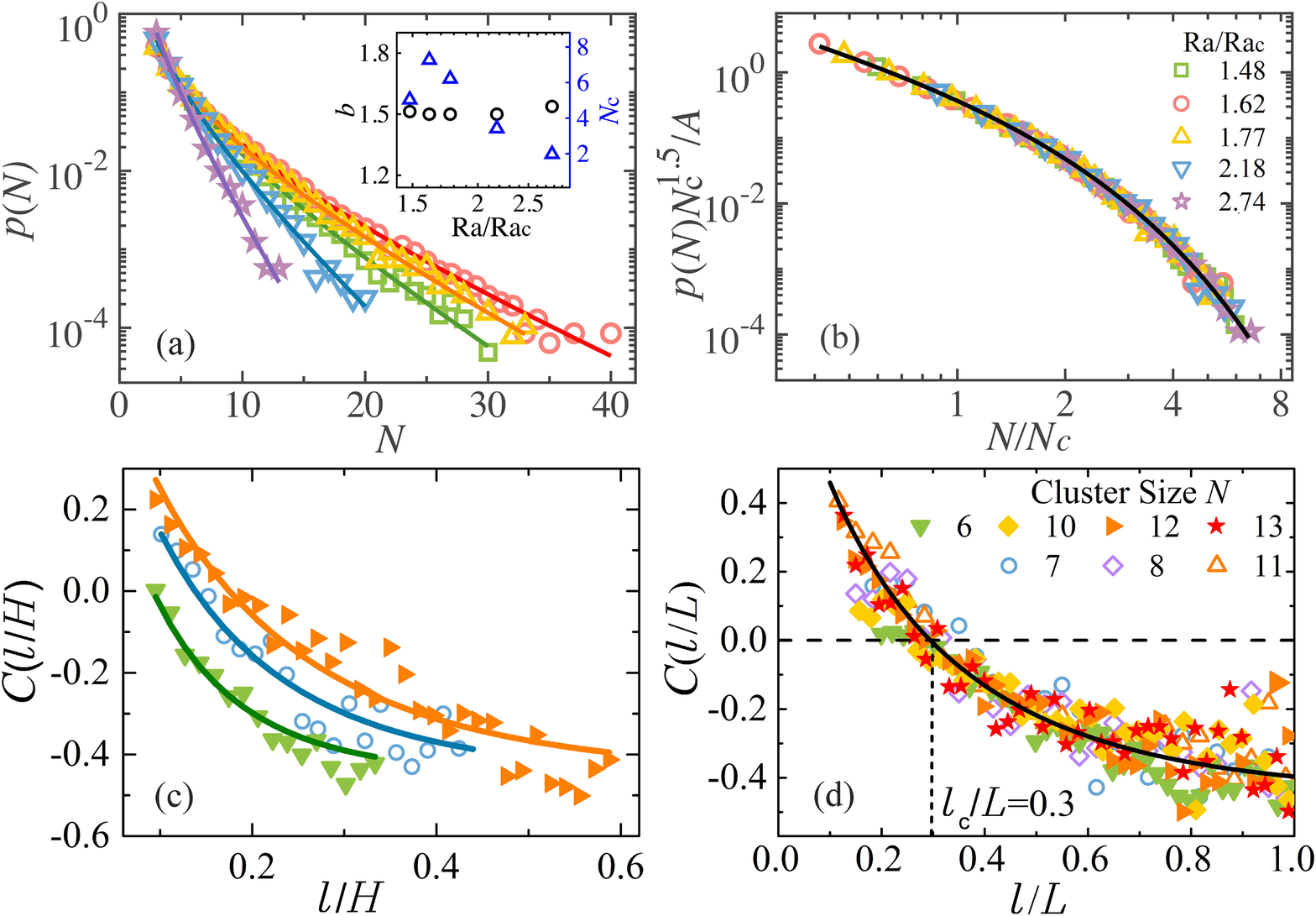}
\caption{(a) Size distribution of vortex clusters. Solid curves represent fits to the experimental data $p(N){=}AN^{-b}e^{-N/N_c}$. Results for $\mathrm{Ra}{=}3.0{\times}10^7$. Symbols are as in (b). Inset: the fitting parameters, i. e., the exponent $b$ (circles) and cutoff size $N_c$ (triangles) as functions of $Ra/Ra_c$. The prefactor $A$ is determined by the normalization relation: $\Sigma_Np(N){=}1$. (b) The rescaled data $p(N){N_c}^{1.5}/A$ shown in a log-log frame as a function of $N/N_c$. All data collapse onto a master curve shown by the solid curve. (c, d) Correlations of the vortex velocity fluctuation as a function of $l/H$ (c), and as a function of $l/L$ (d). Symbols are defined in (d). Results for $\mathrm{Ra}{=}3.0{\times}10^7$, $\mathrm{Ra}/\mathrm{Ra_c}{=}1.62$. Solid (open) symbols: experimental (DNS) data. The solid curves represent stretched exponential functions, $C(l){=}(1{+}a)e^{({-}c_1l)^{c_2}}{-}a$, fitted to the data (described in Supplementary Information 5). The vertical dashed line in (d) indicates the correlation length $l_c{\approx}0.3L$ determined by the zero-crossing position of $C(l/L)$. }
\label{fig5}
\end{figure*}


Dynamical systems consisting of clustered entities often exhibit scale-invariant, collective motions \cite{KFSH17}. Here we further analyze the spatial correlation function of vortex velocity fluctuation within a vortex cluster $C(l){=}\sum_{ij}[{\vec {u'}_i(\vec r_i{+}\vec l){\cdot}\vec {u'}_j (\vec r_j) \delta(l{-}l_{ij})}]/[C_0{\cdot}\sum_{ij}\delta(l{-}l_{ij})]$, where $\vec u'_i{=}\vec{u}_i{-}\vec{V}$ is the relative vortex velocity around the mean cluster velocity $\vec{V}{=}\sum_i{\vec {u}_i}/N$, $l_{ij}$ is the distance between the vortex pair $(i, j)$ and $C_0$ is a normalization constant. Figure 5c shows that $C(l)$ decreases as the distance $l$ increases, with the decay length depending on the cluster size $N$. In Fig.\ 5d we present the correlation function $C(l/L)$, scaled by the cluster length $L$, for clusters with various sizes. This rescaling leads to the converging of the data onto a single curve representing a stretched exponential function, which crosses zero at the correlation length $l_c{\approx}0.3L$ for all cluster sizes $N$. Thus the correlated motions of the vortices are long-range and scale-free, i.e. there is no characteristic length scale here except the length $L$ of the cluster. We remark that the scattering of data points at large distances ($l{\approx}L$), owing to insufficient statistics, has negligible influence in the determination of $l_c$.

\section*{Discussions}


We have shown the formation of a novel type of large-scale coherent structures, in the form of vortex clusters, in rotating turbulent convection. Within each cluster the lighter cyclones submit to collective motions dominated by the denser anticyclones, exhibiting inverse-centrifugation, outward motion. We find that the size-distribution $p(N)$ of the vortex clusters can be well represented by a fractional power function with an exponential cutoff. The observed robust three-half power scaling of $p(N)$ for small $N$ (see Supplementary Fig.\ 3) suggests that the theory of aggregation \cite{TNT88, Ta89, BDF99} apply to a broad range of grouping phenomena, and may provide predictions for the clustering dynamics of vortices in the present highly nonlinear, turbulent systems. 

For large $N$ we find that $p(N)$ decays exponentially and the cutoff size $N_c$ is maximum when the centrifugal effect is dominant. 
Further investigations reveal that $N_c$ is proportional to the ratio of the vortex population density over the splitting rate of the vortices from the clusters (detailed in Supplementary Fig.\ 4), analogous to various biological systems \cite{BDF99, Ni03}. We thus attribute the exponential decay of $p(N)$ to the splitting and aggregating process of vortices between the clusters and the ambient flows, which maintains a statistically stable cluster-size distribution. As is shown by Fig.\ 4 in Supplementary Information, the vortex splitting rate reaches a minimum when the asymmetry of the vorticity fields is maximum. Thus when the anticyclonic flows dominate mostly the interaction between adjacent vortices, the cluster structures possess the maximum stability against splitting, leading to the largest characterized cluster size $N_c$.     

Last, we discover that the self-organized vortices exhibit scale-free correlations of velocity fluctuations, with the correlation length being approximately $30\%$ cluster length. This phenomenon of scale-invariant dynamics is analogous to the collective behavior observed widely in bird flocks, bacterial colony \cite{CCGPSSV10, CDBSZ12} and in active matters \cite{KFSH17}. The present study brings new perspectives on the phenomenon of collective motion, and may have broad implications in the studies of soft condensed matter, fluid physics and biological systems.

\section*{Acknowledgements}
This work is supported by the National Science Foundation of China under Grant No. 11572230 and 11772235, a NSFC/RGC Joint Research Grant No. 11561161004 (JQZ) and N\_CUHK437$/$15 (KQX) and by the Hong Kong Research Grants Council under Grant No. 14301115 and 14302317, and by a SUSTech Startup Fund.

\section*{Author contributions}
J.-Q.Z. and K.-Q.X. conceived and designed research. S-S.D., J.-Q.S. and H.-Y.L. conducted the experiments. K.L.C and G.-Y.D. conducted the numerical simulations. J.-Q.Z. and K.-Q.X. wrote the manuscript. S.-S.D and K.L.C. contributed equally to this study.

\section*{Additional information}
Supplementary information is available in the online version of the paper. Reprints and permissions information is available online. Correspondence and requests for materials should be addressed to J.-Q. Z. 

\section*{Competing financial interest}
The authors declare no competing financial interests.

\section*{Methods}

\textbf{Experimental set-up.}
The experimental apparatus had been used for several previous investigations of turbulent rotating RB convection \cite{Paper1, SLDZ20}. In the present study we used a cylindrical cell mounted on a rapidly rotating table. Its bottom plate was made of 35 mm thick oxygen-free copper, heated from below by a uniformly-distributed electric wire heater. Its top plate was a 5 mm thick sapphire disc, cooled from above through circulating coolant. Its sidewall, made of 3mm thick Plexiglas, was protected against the ambient temperature fluctuations by an adiabatic shield that maintained a constant temperature. For flow visualization, a PIV system was installed on the co-rotating frame. A thin light-sheet powered by a solid-state laser illuminated the seed particles in a horizontal plane at a fluid height $z{=}H/4$ (Fig.\ 1i). Images of the particle were captured through the top sapphire window by a high-resolution camera (2448$\times$2050 pixels). Two-dimensional velocity fields were extracted by cross-correlating two consecutive particle images. Each velocity vector was calculated from an interrogation windows (32$\times$32 pixels), with $50\%$ overlap of neighboring sub-windows to ensure sufficient accuracy and resolution \cite{WEA13}. For each measurement, we took image sequences at a time interval of 0.5 sec with typical acquisition time of 2.5 hrs.  Detailed experimental schemes of vortex identification and tracking are provided in Supplementary Information 1. \\

{\noindent}\textbf{Numerical method.}
In the DNS we solved the three-dimensional Navier-Stokes equations with the Boussinesq approximation: 
\begin{equation} 
\begin{split}
\frac{D\vec{u}}{Dt}=&-{\nabla}P+(\frac{\mathrm{Pr}}{\mathrm{Ra}})^{1/2}{\nabla}^2\vec{u}+{\theta}\hat{z} \\
&+(\frac{\mathrm{Pr}}{\mathrm{RaEk}^2})^{1/2}\vec{u}{\times}\hat{z}-\frac{2r\mathrm{Fr}}{d}{\theta}\hat{r}, 
\end{split}
\end{equation}
\begin{equation} 
\frac{D{\theta}}{Dt}=\frac{1}{(\mathrm{RaPr})^{1/2}}{\nabla}^2{\theta},
\end{equation}
\begin{equation} 
{\nabla}{\cdot}\vec{u}=0.
\end{equation}
Here $\vec{u}$ is the fluid velocity, $\theta$ and $P$ are the reduced temperature and pressure. The last two terms in the momentum equation (Eq.\ 1) represent the Coriolis force and the centrifugal force. The equations were nondimensionalized using $L$, ${\Delta}T$ and the free-fall velocity $U_{f}{=}\sqrt{\alpha g\Delta T L}$. The simulations were performed in a cylindrical sample with an aspect ratio $\Gamma{=}4$ and no-slip boundaries at all walls. Equations 1-3 were solved using a fully parallelized direct numerical simulation code {\it{CUPS}} based on finite volume method with 4th order precision \cite{KX13, CDX18}. To increase computational efficiency without any sacrifice in precision, we used a multiple-resolution strategy, in which the temperature equation was solved in a finer grid than the momentum one, allowing for a sufficient resolution to resolve the Batchelor and Kolmogorov length scales \cite{CDX18}. In addition, we considered the flow fields with $Fr{=}0$ \cite{Paper1}. In this case the governing equations were solved under Cartesian coordinates with periodic boundary conditions.


\begin{thebibliography}{10}

\bibitem{Va06}
Vallis{,} G.~K. 
\newblock {\em Atmospheric and Oceanic Fluid Dynamics}.
\newblock (Cambridge Univ. Press, Cambridge, 2006).

\bibitem{Mc06}
McWilliams{,} J.~C. 
\newblock {\em Fundamentals of Geophysical Fluid Dynamics}.
\newblock (Cambridge Univ. Press, Cambridge, 2006).

\bibitem{HV93}
Hopfinger{,} E.~J. \& van Heijst{,} G.~J.~F. 
\newblock Vortices in rotating fluids.
\newblock {\em Annu. Rev. Fluid Mech.} \textbf{25}, 241--289 (1993).

\bibitem{VK89}
van Heijst{,} G.~J.~F. \& Kloosterziel{,} R.~C. 
\newblock Tripolar vortices in a rotating fluid.
\newblock {\em Nature}  \textbf{338}, 569--571 (1989).

\bibitem{AMetal18}
Adriani{,} A. {\em et~al}.
\newblock Clusters of cyclones encircling jupiter`s poles.
\newblock {\em Nature} \textbf{555}, 216 (2018).

\bibitem{YB20}
Yadav{,} R.~K. \& Bloxham{,} J.
\newblock Deep rotating convection generates the polar hexagon on saturn.
\newblock {\em Proc. Natl. Acad. Sci. USA} \textbf{117}, 13991--13996 (2020).

\bibitem{LM93}
Legg{,} S. \& Marshall{,} J.
\newblock A heton model of the spreading phase of open-ocean deep convection.
\newblock {\em J. Phys. Oceanogr.} \textbf{23}, 1040--1056 (1993).

\bibitem{BW99}
Bush{,} J.~W.~M. \& Woods{,} A.~W. 
\newblock Vortex generation by line plumes in a rotating stratified fluid.
\newblock {\em J. Fluid Mech.} \textbf{388}, 289--313 (1999).

\bibitem{FS01}
Fernando{,} H.~J.~S. \& Smith{,} D.~C. 
\newblock Vortex structures in geophysical convection.
\newblock {\em Eur. J. Mech. B - Fluids} \textbf{20}, 437--470 (2001).

\bibitem{CCGPSSV10}
Cavagna{,} A. {\em et~al}.
\newblock Scale-free correlations in starling flocks.
\newblock {\em Proc. Natl. Acad. Sci. USA} \textbf{107}, 11865--11870 (2010).

\bibitem{CDBSZ12}
Chen{,} X., Dong{,} X., Be'er{,} A., Swinney{,} H.~L. \& Zhang{,} H.~P. 
\newblock Scale-invariant correlations in dynamic bacterial clusters.
\newblock {\em Phys. Rev. Lett.}  \textbf{108}, 148101 (2012).

\bibitem{KFSH17}
Khaluf{,} Y., Ferrante{,} E., Simoens{,} P. \& Huepe{,} C.
\newblock Scale invariance in natural and artificial collective systems: a
  review.
\newblock {\em J. R. Soc. Interface} \textbf{14}, 20170662 (2017).

\bibitem{AGL09}
Ahlers{,} G., Grossmann{,} S. \& Lohse{,} D.
\newblock Heat transfer and large scale dynamics in turbulent
  {{Rayleigh-B\'enard}} convection.
\newblock {\em Rev. Mod. Phys.} \textbf{81}, 503--537 (2009).

\bibitem{LX10}
Lohse{,} D. \& Xia{,} K.-Q.
\newblock Small-scale properties of turbulent {{Rayleigh-B\'enard}} convection.
\newblock {\em Annu. Rev. Fluid Mech.} \textbf{42}, 335-364 (2010).

\bibitem{Xi13}
Xia{,} K.-Q. 
\newblock Current trends and future directions in turbulent thermal convection.
\newblock {\em Theor. Appl. Mech. Lett.} \textbf{3}, 052001 (2013).

\bibitem{LE97}
Liu{,} Y. \& Ecke{,} R.~E. 
\newblock Heat transport scaling in turbulent {{Rayleigh-B\'enard}} convection:
  effects of rotation and {{Prandtl}} number.
\newblock {\em Phys. Rev. Lett.} \textbf{79}, 2257--2260 (1997).

\bibitem{KCG06}
Kunnen{,} R.~P.~J., Clercx{,} H.~J.~H. \& Geurts{,} B.~J. 
\newblock Heat flux intensification by vortical flow localization in rotating
  convection.
\newblock {\em Phys. Rev. E} \textbf{74}, 056306 (2006).

\bibitem{ZSCVLA09}
Zhong{,} J.-Q. {\em et~al}.
\newblock Prandtl-, {{Rayleigh}}-, and {{Rossby}}-number dependence of heat
  transport in turbulent rotating {{Rayleigh-B\'enard}} convection.
\newblock {\em Phys. Rev. Lett.}  \textbf{102}, 044502 (2009).

\bibitem{HA18}
Horn{,} S. \& Aurnou{,} J.~M.
\newblock Regimes of coriolis-centrifugal convection.
\newblock {\em Phys. Rev. Lett.} \textbf{120}, 204502 (2018).

\bibitem{Paper1}
Chong{,} K.~L.{\em et~al}.
\newblock Vortices as {{Brownian}} particles in turbulent flows.
\newblock {\em Sci. Adv.} \textbf{6}, eaaz1110 (2020).

\bibitem{BG86}
Boubnov{,} B.~M. \& Golitsyn{,} G.~S. 
\newblock Experimental study of convective structures in rotating fluids.
\newblock {\em J. Fluid Mech.} \textbf{167}, 503--531 (1986).

\bibitem{Sa97}
Sakai{,} S.
\newblock The horizontal scale of rotating convection in the geostrophic
  regime.
\newblock {\em J. Fluid Mech.} \textbf{333}, 85--95 (1997).

\bibitem{VE02}
Vorobieff{,} P. \& Ecke{,} R.~E. 
\newblock Turbulent rotating convection: an experimental study.
\newblock {\em J. Fluid Mech.} \textbf{458}, 191--218 (2002).

\bibitem{PKVM08}
Portegies{,} J.~W., Kunnen{,} R.~P.~J., van Heijst{,} G.~J.~F. \& Molenaar{,} J.
\newblock A model for vortical plumes in rotating convection.
\newblock {\em Phys. Fluids} \textbf{20}, 066602 (2008).

\bibitem{KSNHA09}
King{,} E.~M., Stellmach{,} S., Noir{,} J., Hansen{,} U. \& Aurnou{,} J.~M. 
\newblock Boundary layer control of rotating convection systems.
\newblock {\em Nature} \textbf{457}, 301--304 (2009).

\bibitem{GJWK10}
Grooms{,} I., Julien{,} K., Weiss{,} J.~B. \& Knobloch{,} E.
\newblock Model of convective {{Taylor}} columns in rotating
  {{Rayleigh-B\'enard}} convection.
\newblock {\em Phys. Rev. Lett.} \textbf{104}, 224501 (2010).

\bibitem{KCG10}
Kunnen{,} R.~P.~J., Clercx{,} H.~J.~H. \& Geurts{,} B.~J. 
\newblock Vortex statistics in turbulent rotating convection.
\newblock {\em Phys. Rev. E} \textbf{82}, 036306 (2010).

\bibitem{JRGK12}
Julien{,} K., Rubio{,} A.~M., Grooms{,} I. \& Knobloch{,} E.
\newblock Statistical and physical balances in low rossby number
  {{Rayleigh-B\'enard}} convection.
\newblock {\em Geophys. Astrophys. Fluid Dyn.} \textbf{106}, 392--428 (2012).

\bibitem{KA12}
King{,} E.~M. \& Aurnou{,} J.~M. 
\newblock Thermal evidence for {{Taylor}} columns in turbulent rotating
  {{Rayleigh-B\'enard}} convection.
\newblock {\em Phys. Rev. E} \textbf{85}, 016313 (2012).

\bibitem{SLDZ20}
Shi{,} J.-Q., Lu{,} H.-Y., Ding{,} S.-S. \& Zhong{,} J.-Q. 
\newblock Fine vortex structure and flow transition to the geostrophic regime
  in rotating {{Rayleigh-B\'enard}} convection.
\newblock {\em Phys. Rev. Fluids} \textbf{5}, 011501 (2020).

\bibitem{Tr88}
Tritton{,} D.~J. 
\newblock {\em Physical Fluid Dynamics}.
\newblock (Oxford Univ. Press, New York, 1988).

\bibitem{Ch61}
Chandrasekhar{,} S.
\newblock {\em Hydrodynamic and Hydromagnetic Stability}.
\newblock (Oxford Univ. Press, Oxford, 1961).

\bibitem{KX13}
Kaczorowski{,} M. \& Xia{,} K.-Q.
\newblock Turbulent flow in the bulk of {{Rayleigh-B\'enard}} convection:
  small-scale properties in a cubic cell.
\newblock {\em J. Fluid Mech.} \textbf{722}, 596--617 (2013).

\bibitem{CDX18}
Chong{,} K.~L., Ding{,} G. \& Xia{,} K.-Q.
\newblock Multiple-resolution scheme in finite-volume code for active or
  passive scalar turbulence.
\newblock {\em J. Comp. Phys.} \textbf{375}, 1045--1058 (2018).

\bibitem{HO99}
Hart{,} J.~E. \& Olsen{,} D.~R. 
\newblock On the thermal offset in turbulent rotating convection.
\newblock {\em Phys. Fluids} \textbf{11}, 2101--2107 (1999).

\bibitem{LE11}
Liu{,} Y. \& Ecke{,} R.~E. 
\newblock Local temperature measurements in turbulent rotating
  {{Rayleigh-B\'enard}} convection.
\newblock {\em Phys. Rev. E.} \textbf{84}, 016311 (2011).

\bibitem{HA19}
Horn{,} S. \& Aurnou{,} J.~M. 
\newblock Rotating convection with centrifugal buoyancy: Numerical predictions
  for laboratory experiments.
\newblock {\em Phys. Rev. Fluids} \textbf{4}, 073501 (2019).

\bibitem{TNT88}
Takayasu{,} H., Nishikawa{,} I. \& Tasaki{,} H.
\newblock Power-law mass-distribution of aggregation systems with injection.
\newblock {\em Phys. Rev. A} \textbf{37}, 3110-3117 (1988).

\bibitem{Ta89}
Takayasu{,} H.,
\newblock Steady-state distribution of generalized aggregation system with injection.
\newblock {\em Phys. Rev. Lett.} \textbf{63}, 2563-2565 (1989).

\bibitem{BDF99}
Bonabeau{,} E., Dagorn{,} L. \& Freon{,} P.
\newblock Scaling in animal group-size distributions.
\newblock {\em Proc. Natl. Acad. Sci. USA} \textbf{96}, 4472--4477 (1999).

\bibitem{LLW16}
Leweke{,} T., Le Diz\'es{,} S. \& Williamson{,} C.~H.~K. 
\newblock Dynamics and instabilities of vortex pairs.
\newblock {\em Annu. Rev. Fluid Mech.} \textbf{48}, 507--541 (2016).

\bibitem{BD95}
Bonabeau{,} E. \& Dagorn{,} L.
\newblock Possible universality in the size distribution of fish schools.
\newblock {\em Phys. Rev. E} \textbf{51}, R5220--R5223 (1995).

\bibitem{Ni03}
Niwa{,} H.-S. 
\newblock Power-law versus exponential distributions of animal group sizes.
\newblock {\em J. Theor. Biol.} \textbf{224}, 451--457 (2003).

\bibitem{WEA13}
Westerweel{,} J., Elsinga{,} G.~E. \& Adrian{,} R.~J. 
\newblock Particle image velocimetry for complex and turbulent flows.
\newblock {\em Annu. Rev. Fluid Mech.} \textbf{45}, 409--436 (2013).

\end{thebibliography}


\end{document}